# Bayesian multi-parameter evidence synthesis to inform decision-making: a case study in hormone-refractory metastatic prostate cancer

Sze Huey Tan[1, 2], Keith R. Abrams[1] and Sylwia Bujkiewicz[1]

[1] Biostatistics Research Group, Department of Health Sciences, University of Leicester, University Road, Leicester, LE1 7RH, UK

[2] Division of Clinical Trials and Epidemiological Sciences, National Cancer Centre Singapore, 11 Hospital Drive, Singapore 169610, Singapore

**Abstract**

In health technology assessment, decisions are based on complex cost-effectiveness models which, to be implemented, require numerous input parameters. When some of the relevant estimates are not available the model may have to be simplified. Multi-parameter evidence synthesis allows to combine data from diverse sources of evidence resulting in obtaining estimates required in clinical decision-making that otherwise may not be available. We demonstrate how bivariate meta-analysis can be used to predict unreported estimate of a treatment effect enabling implementation of multi-state Markov model, which otherwise needs to be simplified. To illustrate this, we used an example of cost-effectiveness analysis for docetaxel in combination with prednisolone in metastatic hormone-refractory prostate cancer (mHRPC). Bivariate meta-analysis was used to model jointly available data on treatment effects on overall survival (OS) and progression-free survival (PFS) to predict the unreported effect on PFS in a study evaluating docetaxel with prednisolone. Predicted treatment effect on PFS allowed implementation of a three-state Markov model comprising of stable disease, progressive disease and death states, whilst lack of the estimate restricted the model to two-state model (with stable disease and death states). The two-state and three-state models were compared by calculating the incremental cost-effectiveness ratios, which was much lower in the three-state model: £21966 per QALY gained compared to £30026 obtained from the two-state model. In contrast to the two-state model, the three-state model has the advantage of distinguishing patients who progressed from those who did not progress. Hence the use of advanced meta-analytic technique helped to obtain relevant parameter estimate to populate a model which describes natural history more accurately, and at the same helped to prevent valuable clinical data from being discarded.

# 1 Introduction

In health technology assessment (HTA), reimbursement decisions for new health technologies are made based on cost-effectiveness models. Such, often complex, models are implemented using estimates of effectiveness, health related quality of life (HRQoL) and cost. Effectiveness estimates are usually obtained from the systematic literature review and meta-analysis of randomised controlled trials (RCTs) which are designed to give an estimate of the treatment effect on the primary clinical outcome. The choice of the outcome measures for RCTs and reporting of findings rarely takes into consideration what is important from the HTA perspective and therefore trials not always report all outcomes relevant to the HTA. There is often a lot of heterogeneity in reporting of clinical outcomes due to, for example, variety of scales on which effectiveness can be measured, different time points at which different studies report their outcomes or different control arms. Relevant outcomes may not be reported due to poor study design, outcome reporting bias or some problems with outcome measurement. This may lead to difficulties with populating cost-effectiveness model with appropriate parameters.

Bayesian statistics provides flexible framework for modelling complex data structures by allowing multiple parameters to be modelled simultaneously. This is particularly useful when multiple data sources need to be brought together, which can be achieved by the use of multi-parameter evidence synthesis. Network meta-analysis (NMA) facilitates simultaneous comparison of multiple treatment options with an aim to obtain effectiveness estimates for all possible treatment contrasts including those that may not be directly reported by any RCTs[1]. Multivariate meta-analysis allows to model jointly treatment effects on multiple outcomes with the aim of obtaining poled effect on all the outcomes while taking into account of the correlation between them[2-4]. There are many advantages of multivariate meta-analysis, including (i) potentially increased precision of effectiveness estimates which can lead to increased precision of other estimates required in decision modelling, such as HRQoL[5],(ii) inclusion of all relevant evidence from studies reporting relevant outcomes (other than the main outcome of interest) preventing valuable data from being discarded and (iii) potentially reduced outcome reporting bias[6]. In this paper, we propose the use of bivariate meta-analysis for purpose of predicting unreported treatment effects in individual studies, rather than obtaining overall pooled effects[4, 7, 8], for purpose of informing a complex HTA modelling framework.

A multi-state Markov model is one of the most frequently used decision models in HTA. The number of health states in the model depends on the specific disease area and the states should be chosen to represent important events (clinically and economically) and be mutually exclusive such as, for example, three-state model including asymptotic (or stable disease) state, progressive disease state and death[9]. To populate a three state the model in cancer, the transition probabilities between the states are obtained from data on both overall survival (OS) and progression free survival (PFS). When data are not available to estimate all parameters of the model, such model may need to be simplified which conflicts with its purpose to simulate real life scenarios. We illustrate how multi-parameter evidence synthesis can be used to fully utilize the available evidence to inform parameters of Markov model.

To demonstrate this methodology, we apply the methods to inform cost-effectiveness analysis of docetaxel in prostate cancer. In the technology assessment of docetaxel, Collins et al. constructed a two-state Markov model consisting of stable disease and death states[10]. Reviewing the evidence on effectiveness of docetaxel made it apparent that relevant evidence on the treatment effect on progression was not available at the submission stage of evidence for docetaxel, limiting the development of the decision model to the two states only. We will demonstrate how the use of multivariate meta-analysis can lead to obtaining relevant estimates necessary to populate the three-state Markov model including a progression state.

## 2. Methods

### *2.1 Motivating example and sources of evidence*

In 2007 the National Institute for Health and Care Excellence (NICE) carried out a technology appraisal of use of docetaxel in combination with either prednisone or prednisolone (D+P) as treatments for metastatic hormone-refractory prostate cancer (mHRPC)[10]. The technology appraisal aimed to evaluate the clinical- and cost-effectiveness of the combination therapy. Evidence base for the meta-analysis in this HTA included four studies which investigated interventions that were licensed at the time of the HTA submission, namely D+P, mitoxantrone plus prednisone (M+P), prednisolone alone (P), mitoxantrone plus hydrocortisone (M+H) and hydrocortisone alone (H). Data from the four RCTs[11-14], listed in Table 1, are included in our example and referred to as HTA set. None of the studies reported the effect of docetaxel on PFS, required to populate three-state Markov model for the cost-effectiveness assessment of this treatment and only one trial reported the effect of docetaxel on OS (TAX 327[13]). The details of the systematic review conducted by Collins et al[10] are included in the Appendix A of Supplementary Materials, which also includes set of studies of unlicensed treatments used to inform some of the model parameters in this paper.

*Table 1: Randomised controlled trials in HTA report base clinical-effectiveness analysis*

| Trial | Year | No. of arms | Reference Treatment | Comparative Treatment(s) | Total no. of patients | OS data | PFS data |
|---|---|---|---|---|---|---|---|
| CCI-NOV22[14] | 1996 | 2 | M+P | P | 161 | Yes | Yes |
| CALGB 9182[12] | 1999 | 2 | M+H | H | 242 | Yes | Yes |
| Berry et al.[11] | 2002 | 2 | M+P | P | 120 | Yes | Yes |
| TAX 327[13] | 2004 | 3 | M+P | D+P<br>D1+P | 1006 | Yes | No |

### 2.2 Methods of evidence synthesis for predicting treatment effect on PFS for docetaxel

Figure 1 shows a schematic diagram of the procedures for the prediction of PFS hazard ratio (HR) of D+P versus M+P as well as the use of indirect comparisons to obtain the estimates of both PFS and OS HRs for D+P versus P. For purpose of conducting indirect comparisons meta-analysis, we group both corticosteroids P and H denoting them as P. The components of the analysis were conducted in the following order:

1. To unify the scale of the treatment effects across studies, Kaplan-Meier curves for OS and PFS were used, where available, to reconstruct the individual patient data (IPD) for the four RCTs.
2. Treatment effects on log HR scale were estimated by performing survival analysis on the reconstructed IPD.
3. A bivariate meta-analytic model was used to model jointly log HRs on PFS and OS resulting in predicted estimate of the log HR on PFS for trial TAX 327 (that compared interventions: D+P to M+P) which was not reported.
4. The predicted log HR on PFS for trial TAX 327 was then used in a random-effect indirect comparison meta-analysis (ICMA) (also conducted for OS analysis in the HTA report) to obtain the estimate of the effect of D+P versus P on PFS.

Methods for obtaining the summary trial data for this analysis on appropriate scale are described in Section 2.2.1. Meta-analytic method for combining evidence on both PFS and OS for purpose of predictions is described in section 2.2.2, whilst section 2.2.3 discusses briefly ICMA.

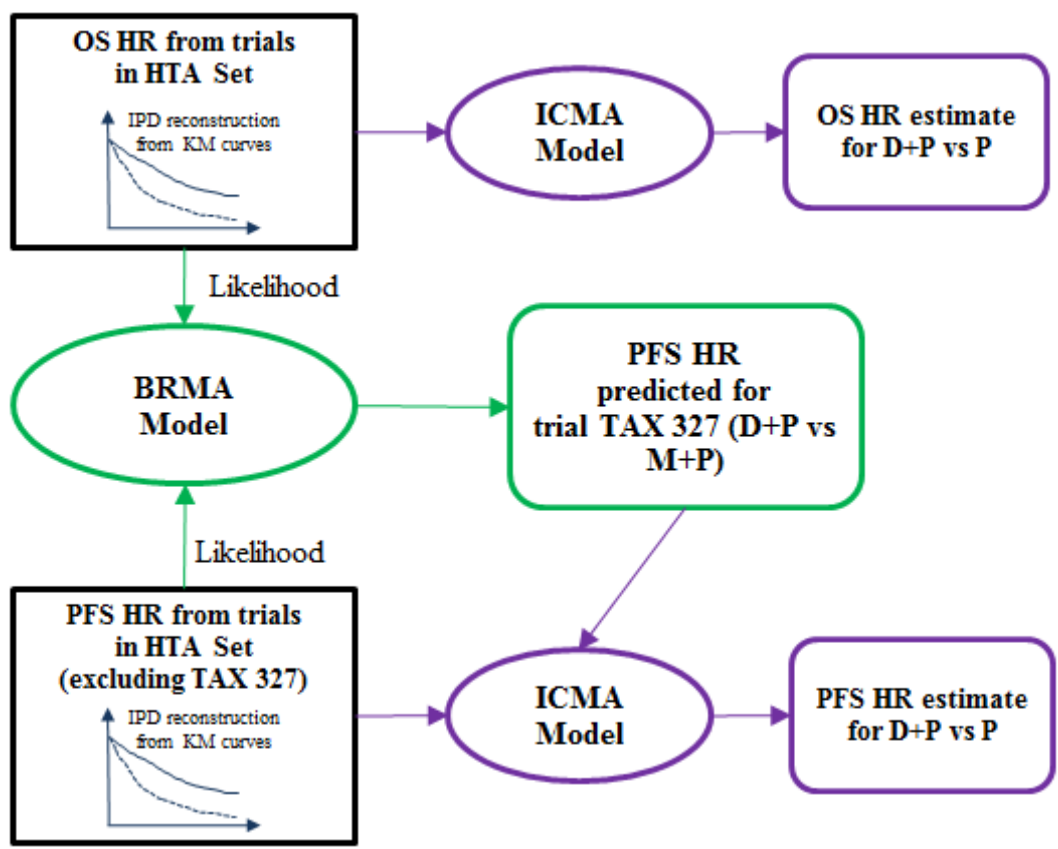

*Figure 1: Diagram for the clinical effectiveness analysis using BRMA*

*2.2.1 Data extraction and reconstruction*

For the purpose of evidence synthesis, summary data on effectiveness measured on OS and PFS were analysed on the log HR scale, to allow for assumption of normality of the effect estimates. To obtain the estimates on this scale, IPD on OS and PFS for each of the RCTs were reconstructed from their respective Kaplan-Meier survival curves, if reported, using the method proposed by Guyot and colleagues[15]. Reconstructed IPD allow log HRs and corresponding standard errors (SEs) to be estimated using survival analysis instead of crude estimation using median survival times and log-rank test p-values reported in the RCTs. Survival analyses, using Cox model, were performed on the reconstructed IPD of the four RCTs in the HTA Set for OS and two in the HTA Set for PFS RCTs (only CALGB9182[12] and Berry[11] reported Kaplan-Meier survival curves for PFS) to estimate the log HRs for the meta-analysis. The estimate of log HR on PFS for trial CCI-NOV22[13] was obtained from the HTA report[10] as it was not reported in the published article. Trial TAX 327 did not report PFS endpoint.

*2.2.2 Bivariate meta-analysis*

Bivariate random-effects meta-analysis (BRMA) was used for purpose of predicting the treatment effect on PFS in the docetaxel study (TAX 327) by modelling treatment effects measured by log HRs on OS and PFS jointly, taking into account of the correlation between them.

In this model, $Y_{OS}$ and $Y_{PFS}$, the treatment effect on OS and PFS on log HR scale, are assumed to be normally distributed and correlated

$$\begin{pmatrix} Y_{OS,i} \\ Y_{PFS,i} \end{pmatrix} \sim Normal\left( \begin{pmatrix} \mu_{OS,i} \\ \mu_{PFS,i} \end{pmatrix}, \Sigma_i \right) \quad (1)$$

with the within-study variance-covariance matrices $\Sigma_i = \begin{pmatrix} \sigma_{OS,i}^2 & \sigma_{OS,i}\sigma_{PFS,i}\rho_{w,i} \\ \sigma_{PFS,i}\sigma_{OS,i}\rho_{w,i} & \sigma_{PFS,i}^2 \end{pmatrix}$ comprising of the within-study standard errors of the estimates, $\sigma_{OS,i}$ and $\sigma_{PFS,i}$, and the within-study correlation $\rho_{w,i}$ between the estimates in each study $i$. The treatment effects $Y_{OS,i}$ and $Y_{PFS,i}$ are the estimates the true effects, $\mu_{OS,i}$ and $\mu_{PFS,i}$, which are also correlated and can be modelled in the form of the product of univariate conditional normal distributions, the product normal formulation[4, 7, 8]:

$$\mu_{OS,i} \sim Normal(\eta_{OS}, \psi_{OS}^2)$$

$$\mu_{PFS,i} | \mu_{OS,i} \sim Normal(\eta_{PFS,i}, \psi_{PFS}^2)$$

$$\eta_{PFS,i} = \lambda_0 + \lambda_1(\mu_{OS,i} - \overline{\mu_{OS,i}}) \quad (2)$$

where $\psi_{OS}^2$ is equal to the between-studies variance $\tau_{OS}^2$ (heterogeneity parameter) of the treatment effects on OS, $\psi_{PFS}^2$ is the conditional between-studies variance of the treatment effects on PFS conditional on the effect on OS, which relates to the heterogeneity parameter $\tau_{PFS}^2$; $\psi_{PFS}^2 = \tau_{PFS}^2 - \lambda_1^2 \tau_{OS}^2$. The slope $\lambda_1 = \rho_b \tau_{PFS}/\tau_{OS}$ and $\rho_b$ is the between-studies correlation. Prior distributions are placed on the between-studies parameters, for example uniform distribution for the correlation $\rho_b \sim Uniform(-1, 1)$ and half normal distributions for the standard deviations; $\tau_{OS,i} \sim HNormal(0, 10^3)$ and $\tau_{PFS,i} \sim HNormal(0, 10^3)$, which give implied prior distributions on $\lambda_1, \psi_{OS}$ and $\psi_{PFS}$ obtained using the above relationships between the parameters. Prior distributions are also placed on other unknown parameters, the intercept $\lambda_0 \sim Normal(0, 10^3)$ and the within-study correlations $\rho_{w,i} \sim Uniform(-1, 1)$. The pooled treatment effects are $HR_{OS} = \exp(\eta_{OS})$ and $HR_{PFS} = \exp(\lambda_0)$. Further assumption about exchangeability of population variances[4] were made in order to predict the standard error corresponding to the missing HR on PFS, comparing D+P versus M+P, in TAX 327.

The missing, unreported effect on PFS in TAX 327 trial is predicted directly from the MCMC simulation. In ordinary approach to multivariate meta-analysis, predicted effects (on outcomes that are not reported) are by-products of the analysis that contribute to the pooled effects[4]. Here we exploit this by using the predicted value directly to inform decision model.

*2.2.3. Indirect meta-analysis and network meta-analysis*

ICMA allows for the comparisons of interventions when there is no head-to-head RCT that compared them directly by evaluating the difference between the interventions through at least one common comparator[16, 17]. This method evaluates the relative effectiveness between the two interventions using two sets of RCTs that compare each of the interventions with the common comparator separately. If, for example, for three interventions A, B, C there is no RCT that directly compares A and B but there are RCTs that compared A with C and B with C, then the relative effect of A versus B can be estimated indirectly as AB$_{indirect}$ = AC$_{direct}$ – BC$_{direct}$ where AC$_{direct}$ and BC$_{direct}$ represent the direct evidence from RCTs of the treatment effect of A versus C and the treatment effect of B versus C respectively. The method extends to NMA combining both direct and indirect evidence from multiple studies of multiple interventions[1]. NMA was used in this paper to obtain HRs for both OS and PFS comparing M+P versus P alone, which is an alternative contrast that can be used in health economic model as for OS by Collins et al[10].

## 2.3 Methods of cost-effectiveness analysis

Collins at al. specified a two-state Markov model, including the stable disease (StD) and death (D) states, that assessed the cost-effectiveness of (D+P) for the treatment of mHRPC[10]. Figure 2 shows the two-state model specified in the HTA report and the three-state model proposed in this paper, which includes a progressive disease (PD) state. For the comparison of the results of the proposed three-state model with the results of the two-state model reported in the HTA, the two-state model was reproduced to ensure that the models (two-state and three-state) are comparable in terms of input parameters and software used (Excel software was used to implement the original HTA model). For clarity, the re-produced two-state model will be called the "WinBUGS two-state model" and the original two-state model in the HTA report will be called the "HTA two-state model" in this paper. Similarly to the HTA two-state model, both the WinBUGS two-state model and WinBUGS three-state model were run for 180 cycles, where one cycle represented one month and the time horizon was 15 years. A cohort size of 10,000 was used in each of the models.

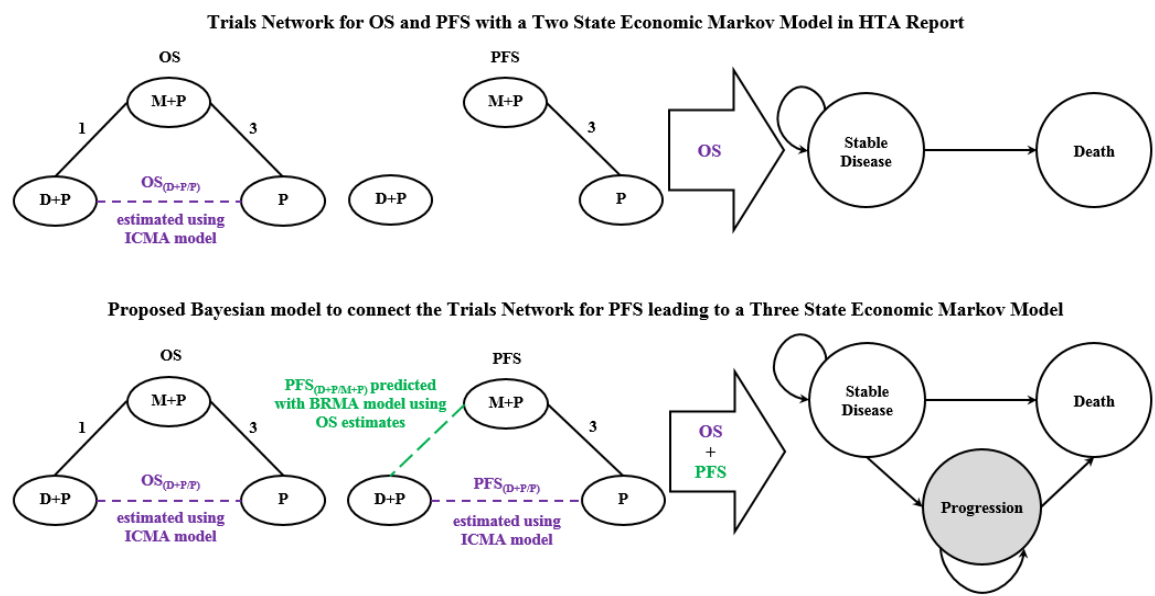

*Figure 2: Original HTA model (top) and proposed Bayesian BRMA to predict PFS for the specification of a three-state economic Markov model (bottom).*

### 2.3.1 Transition probabilities

For the WinBUGS two-state model, the transition probabilities were estimated using the Weibull parameters reported in the HTA report, obtained from trial TAX 327. As for the three-state model, which incorporated a PD state, the transition probabilities for transition from StD state to PD state were estimated using parametric Weibull survival modelling on reconstructed PFS IPD from SWOG trial, which was one of the RCTs of unlicensed drugs used in the sensitivity analysis in the HTA report. It was most comparable trial with TAX 327 (for details about the studies see Appendix A and the justification of the choice of the SWOG trial Appendix C.2 of the supplement). This transition probability from StD state to PD state was obtained for M+P. Transition probability for D+P was in turn calculated by applying the predicted PFS HR of D+P versus M+P to the transition probability of M+P. Transition probability for transition from the StD state to death state was obtained from an article on cost-effectiveness analysis for advance hormone-dependent prostate cancer [18] and was applied to the model with no uncertainty as 0.005.

Survival analysis using the parametric Weibull model was used to implement time-dependency in the transition probabilities in the economic models (transition probabilities for transition from StD state to death state in WinBUGS two-state model; and for transition from StD state to PD state in three-state model). Details of this analysis are included in Appendix B.1.1 of the supplement.

Although IPD were reconstructed for (PFS and OS) for the trial selected for estimating the transition probability from StD state to PD state, the reconstructed IPD for PFS and OS were not paired by patient. Hence, it would not be possible to estimate the transition probabilities from PD state to death state using parametric survival analysis performed using reconstructed IPD as was the case for transition probability for StD state to PD state. To overcome this issue, transition probabilities were estimated by assuming the mean total survival time was equal to the weighted sum of combined survival time from stable disease to progression and then to death and the survival time from stable disease directly to death when death occurred from other causes. The analysis is described in more detail in Appendix B.1.2 of the supplement.

Once the transition probabilities were obtained for the baseline treatment (here M+P), the HRs were applied to them to obtain the transition probabilities for the other two interventions (D+P and P alone). For the two-state model, these are the HRs for OS obtained from the HTA report. For the three-state model, the predicted HR on PFS from BRMA was used to obtain the transition probabilities from StD state to PD state for D+P intervention, whilst the HR obtained from the meta-analysis of the three trials evaluating the corticosteroids P(H) vs M+P(H) was used to obtain the transition probability from StD to PD state for P.

### 2.3.2  Cost

Cost data comprises drug acquisition and administration cost for each intervention, cost of the management of adverse side effects and subsequent follow up cost that included cost of further chemotherapy after disease progression, management of side-effects and palliative care cost. Cost for each of the interventions used in the WinBUGS 2-state and 3-state models were extracted from cost data presented in the HTA report. In the report, costs were categorised into three components: namely, (i) the drug acquisition and administration cost, (ii) the follow up cost and (iii) the terminal care cost. In the three-state model, the follow up costs were divided into portions corresponding to StD state and PD state by taking into account that costs of subsequent chemotherapy and hospitalisations accounted for between 70% and 80% of follow-up costs which most likely occurred post-progression and the remaining follow-up cost (20% to 30%) were related to side effects likely to occur prior to progression (but may also be associated with the subsequent chemotherapy post-progression). The annual discount rate of 3.5% was applied from cycle 13 onwards. Details of cost analysis are includes in Appendix B.2 of the supplement.

### 2.3.3.  Quality-adjusted life-years

Quality-adjusted life-years (QALY) were used as a measure of effect in the cost-effectiveness analysis. To estimate the QALY, utility data in the form of HRQoL were required to quantify the health status of patients with mHRPC, as well as the impact the interventions had on the HRQoL (in terms of disease progression and serious adverse effects).

Quality of life data used in the HTA two-state model by Collins et al[10] were extracted from a study conducted by Sandblom and colleagues[19]. The study was appropriate as it reported HRQoL values using a generic HRQoL instrument, the EuroQoL five-dimensional (EQ-5D) questionnaire that is required in submissions for technology assessment by NICE, it used the population representative of the target population of the HTA; and it provided end-of-life HRQoL values of prostate cancer patients in their last year before death. The EQ-5D values used in the HTA two-state model were also used in our WinBUGS two-state model as well as in the three-state model.

With the inclusion of a PD state in the three-state model, additional utility data for the PD state were required. EQ-5D value used in the two-state models for the StD state were used to describe the HRQoL of patients in the PD state (denoted as $U_{PD}$) in the three state model. This was based on the argument that this utility reflects the patients' HRQoL prior to prostate cancer death and the state prior to death would be the PD state. As the EQ-5D value for the StD state in the two-state model was used to describe the utility for progressed patients, utility for the StD state in the three-state model had to be estimated. This is achieved by splitting the patients in the StD state into three groups and using the following EQ-5D values from the Sandblom et al. study: (i) EQ-5D values of all patients who died of other causes (denoted as $U_{Other\ causes}$) (ii) EQ-5D values of all patients who were still surviving with prostate cancer (denoted as $U_{Surviving}$) and lastly (iii) $U_{PD}$ described earlier. These EQ-5D values, weighted by the transition probabilities amount to the utility for the StD state:

$$U_{SD} = TP_{StD}U_{Surviving} + TP_{StDtoPD}U_{PD} + TP_{StDtoDeath}U_{Other\ causes}$$

where $TP_{StD}$ is the probability of remaining in StD state, $TP_{StDtoPD}$ the probability of transition from StD state to PD state and $TP_{StDtoDeath}$ the probability of transition from StD state to death state without progression. A utility value of zero was assigned to the death state ($U_{Death} = 0$). The utilities for the other states are given in Appendix B.3 of the supplement. Similarly as for cost, an annual discount rate of 3.5% was used for discounting the utilities after the first year.

*2.3.4. Cost-effectiveness analysis*

For the assessment of the cost-effectiveness of the interventions in each model, the mean costs and mean QALYs gained for the interventions and the incremental cost-effectiveness ratios (ICERs) for the comparison of the two interventions of interest (M+P and D+P) were estimated. Cost effectiveness acceptability curves (CEACs) were generated to compare the three interventions. The CEAC and the cost-effectiveness plane were used to compare the proposed three-state model with the WinBUGS two-state model (which is expected to be comparable to the HTA two-state model) when evaluating the difference between D+P and M+P.

### *2.4. Software implementation*

IPD were reconstructed from the Kaplan-Meier curves using the DigitizeIt[20] and R[21]. Survival analyses were implemented in Stata[22]. BRMA and the cost-effectiveness models were implemented using Markov Chain Monte Carlo (MCMC) simulations in WinBUGS[23, 24], with 30, 000 MCMC iterations and 15,000 burn-in (iterations that were discarded) for BRMA and 50,000 iterations and 30,000 burn-in for the cost-effectiveness models. Output data were processed using R[21].

### 3. Results

### *3.1 Clinical effectiveness*

Kaplan-Meier curves for OS were reported for the four trials in the HTA set. Progression-free survival was not reported for TAX 327 and PFS Kaplan-Meier curve was not reported for CCI-NOV22. Hazard ratios, for individual studies, calculated using the reconstructed IPD were comparable with the results reported in the original trials' publications. HRs on OS and PFS reported in the original articles and those obtained from the survival analysis of reconstructed IPD are presented in Appendix C of the supplement (Table C1 and C2).

Table 2 presents all HRs for OS and PFS, obtained from the meta-analyses, including those reported in HTA report and obtained by synthesising HRs from reconstructed IPD. Summary estimates for the HR of OS comparing M+P with P, obtained by combining estimates from survival analysis of reconstructed IPD from the three RCTs that directly compared M+P(H) and P(H) (trials in HTA Set excluding TAX 327) were 0.903 (95% CrI 0.751, 1.084) and 0.901 (95% CrI 0.405, 2.023) using fixed-effect and random-effects meta-analysis respectively. The estimates differed slightly from those in the HTA report, which were 0.99 (95% CI 0.82, 1.20) for both fixed-effect and random-effects. The difference in the HR point estimates was largely due to the lower HRs obtained using the reconstructed IPD for trials: CCI-NOV22, CALGB 9182 and Berry, compared to the HRs reported in the HTA report. However, the 95% credible interval (CrI) estimated using fixed-effect meta-analysis was comparable to the 95% CI reported in the HTA report. Hazard ratios comparing the effect of

D+P versus P(H) on OS were 0.688 (95% CrI 0.523, 0.907) and 0.688 (95% CrI 0.300, 1.604) using fixed-effect and random-effects ICMA respectively as compared to the HR of 0.75 (95% CI 0.57, 0.99) for random-effects ICMA published in the HTA report. Similarly as for comparison of M+P(H) vs. P(H), the estimates of HRs obtained from the reconstructed IPD were lower than those reported in the HTA report.

*Table 2: OS and PFS HRs estimated from traditional and indirect comparison meta-analysis using reconstructed IPD*

| Evidence Synthesis | Hazard Ratio (95% Confidence/CrI) | | |
|---|---|---|---|
| | Overall Survival | | Progression-free Survival |
| | Reported in HTA report | Estimated using reconstructed IPD | Estimated using reconstructed IPD |
| Meta-analysis (M+P/P) | | | |
|    Fixed Effect Analysis | 0.99 (0.82, 1.20) | 0.903 (0.751, 1.084) | 0.641 (0.532, 0.772) |
|    Random Effects Analysis | 0.99 (0.82, 1.20) | 0.901 (0.405, 2.023) | 0.619 (0.170, 2.048) |
| Relative estimate (D+P/M+P) | 0.76 (0.62, 0.94)^ | 0.76 (0.620, 0.936)^ | 0.618 (0.383, 0.941)* |
| Indirect comparison (D+P(H)/P(H)) | | | |
|    Fixed Effect Analysis | Not performed | 0.688 (0.523, 0.907) | 0.396 (0.307, 0.512) |
|    Random Effects Analysis | 0.75 (0.57, 0.99) | 0.688 (0.300, 1.604) | 0.381 (0.107, 1.280) |

*^HRs estimated using TAX 327 trial IPD / re-constructed IPD*
*\*HR predicted using BRMA model*

HR for PFS comparing M+P(H) versus P(H), obtained from fixed-effect and random-effects meta-analyses of estimates from reconstructed IPD, were 0.641 (95% CrI 0.532, 0.772) and 0.619 (95% CrI 0.170, 2.048) respectively. No summary estimate for this comparison was reported in the HTA report. Bayesian BRMA model was used to jointly model the correlated treatment effects on OS and PFS to enable the prediction of treatment effect of D+P versus M+P on PFS for trial TAX 327. The predicted PFS HR for the comparison of D+P versus M+P (for TAX 327) was 0.618 (95% CrI 0.383, 0.941). HRs comparing the effect of D+P versus P(H) on PFS were 0.396 (95% CrI 0.307, 0.512) and 0.381 (95% CrI 0.107, 1.280) from fixed-effect and random-effects ICMA respectively.

### *3.2 Cost effectiveness*

Table 3 presents the differences in cost and QALY and ICERs for D+P compared to M+P, for all three models. The details of the costs of interventions and the mean QALY per patient for each of the interventions in the economic models are presented in Appendix C.3. Results showed that the ICER obtained from the proposed three-state model, using a predicted PFS HR of 0.618 (95% CrI: 0.383 to 0.941) for D+P versus M+P was £21966 compared to £30026 obtained from the WinBUGS two-state model (£32706 in the HTA report). Hence, by implementing the three-state model and taking into account the cost and QALY in the PD state, the ICER estimated was lower than that of the HTA (or WinBUGS) two-state model.

*Table 3: Clinical and cost effectiveness of the HTA two-state model and proposed three-state model*

|  | HTA two-state model | WinBUGS two-state model | WinBUGS three-state model |
|---|---|---|---|
| <u>Clinical Effectiveness</u> | | | |
| Predicted PFS HR (D+P/M+P) | Not applicable in HTA report | Not required | 0.618 (0.383, 0.941) |
| <u>Cost-Effectiveness</u> | | | |
| Difference in Cost, Mean (SE) | 5049 | 4624 (4407.83) | 5350 (4243.95) |
| Difference in QALY, Mean (SE) | 0.15437 | 0.154 (0.0676) | 0.244 (0.0638) |
| ICER | 32706 | 30026 | 21966 |

Differences were calculated as estimate for (D+P) minus estimate for (M+P)

The net benefit for each intervention and the probability that each intervention was cost-effective at £20000 and £30000 are presented in Table 4. At £20000, for the two-state models, M+P had the highest probability of being cost-effective amongst all three interventions (but the same as for P intervention in the HTA model) while D+P had the highest probability of being cost-effective in the three-state model. However, at £30000, D+P had the highest probability of being cost-effective amongst all three interventions for all the models, with the probability being as high as 0.52 in the three-state model compared to 0.44 in the two-state model. Figure 3 shows cost-effectiveness acceptability curves, over a range of willingness to pay thresholds, comparing all the interventions using the WinBUGS two-state model and WinBUGS three-state model. At £20000, M+P had the highest probability cost-effective in the two-state model while D+P had the highest probability cost-effective in the three-state model. However, at £30000, D+P had the highest probability of being cost-effective amongst all three interventions in both models.

*Table 4: Net benefit and probability cost-effectiveness table*

| Intervention | Probability cost-effectiveness | | Net Benefit (95% Credible Interval) | |
|---|---|---|---|---|
| | £20,000 | £30,000 | £20,000 | £30,000 |
| *HTA two-state model* | | | | |
| P | 0.39 | 0.33 | Not Reported | Not Reported |
| M+P | 0.39 | 0.29 | Not Reported | Not Reported |
| D+P (3 weekly) | 0.22 | 0.38 | Not Reported | Not Reported |
| *WinBUGS two-state model* | | | | |
| P | 0.32 | 0.28 | 4417 (-5119, 12360) | 12512 (1648, 22500) |
| M+P | 0.36 | 0.28 | 5023 (-2327, 11680) | 13153 (4116, 21800) |
| D+P (3 weekly) | 0.32 | 0.44 | 3479 (-5749, 12560) | 13149 (1859, 24430) |
| *WinBUGS three-state model* | | | | |
| P | 0.31 | 0.27 | 6829 (-4364, 19720) | 15320 (830, 33820) |
| M+P | 0.34 | 0.21 | 7810 (1724, 13090) | 16704 (9635, 23240) |
| D+P (3 weekly) | 0.36 | 0.52 | 7331 (-790, 15240) | 18660 (9394, 27860) |

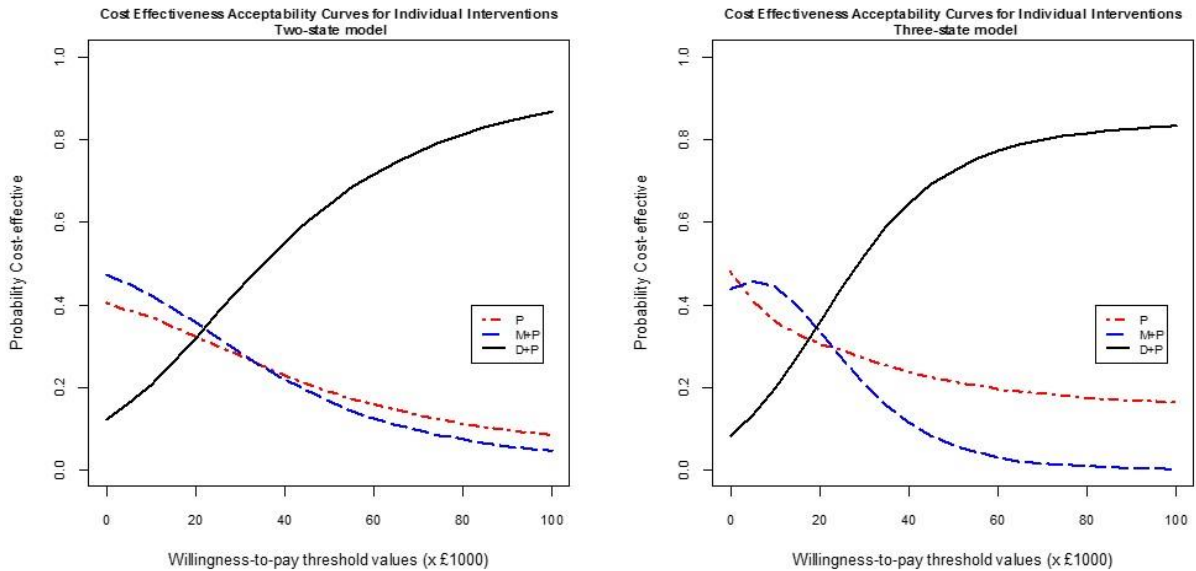

*Figure 3: Cost-effectiveness acceptability curves for all three interventions – two-state (left) and three-state (right) model using direct HR of P versus M+P*

Figure 4 shows cost-effectiveness acceptability curves and cost-effectiveness plane for comparison of the WinBUGS two-state model with the three-state model (using comparison of D+P with M+P). The cost-effectiveness acceptability curves showed that after approximately £10000 willingness-to-pay threshold, the probability that D+P was more cost-effective than M+P was higher in the three-state model than the two-state model. For example, at £30000, the probability was around 0.7 for the three-state model compared to 0.5 for the two-state model.

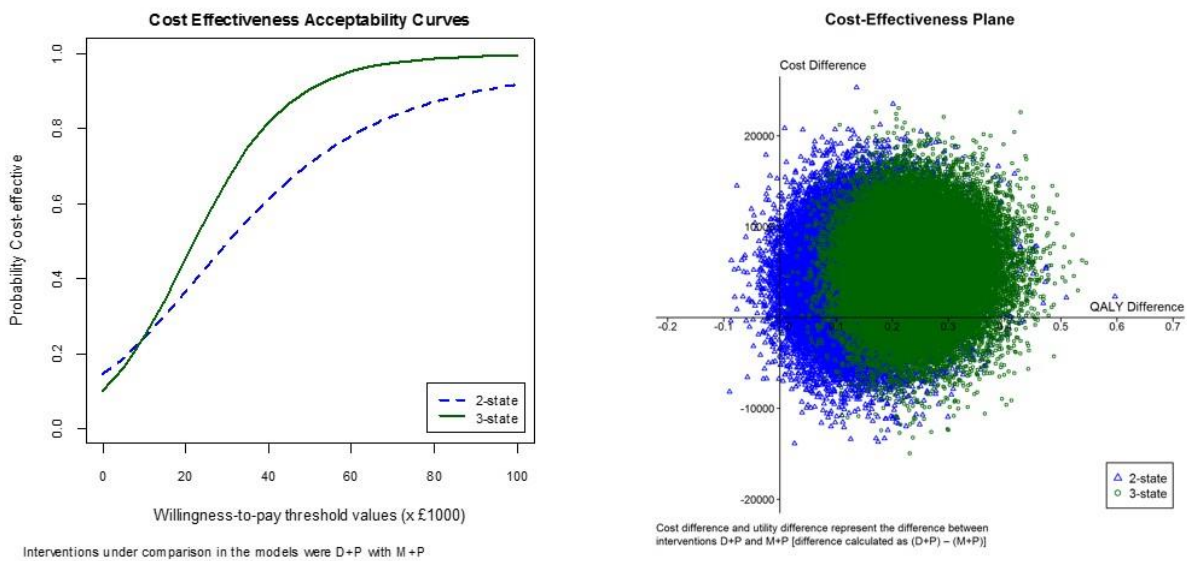

*Figure 4: Cost-effectiveness acceptability curves (left) and cost-effectiveness plane (right) for WinBUGS two-state and three-state economic models*

The cost-effectiveness plane showed that D+P was more effective (in terms of utility) than M+P for both the two-state and three-state models, although the certainty of its effectiveness was greater in the three-state model than the two-state model as all the 50000 Monte Carlo simulations points of the three-state model (green circles) were in the positive utility difference area. Besides being more effective, D+P was also more costly than M+P. However, the degree of uncertainty for the difference in cost of the two interventions were comparable for both models.

## 4. Discussion

We have investigated the use of multi-parameter evidence synthesis, and in particular bivariate meta-analysis, for purpose of allowing all available evidence to inform health economic decision model. The methodology was applied to motivating example of cost-effectiveness of docetaxel (D+P) in mHRPC. We reviewed the original HTA conducted by Collins et al, who implemented two-state Markov model for purpose of cost-effectiveness analysis. The model consisted of stable disease and death states, where the stable disease included both the patients who were stable and those who progressed. The clinical effectiveness estimates used in the model were based on data from the systematic review of RCTs which did not contain any estimate of the effect of docetaxel (in any combination) on PFS, only an estimate of the treatment effect on OS. The aim of our research was to investigate how inclusion of a predicted estimate of the effect of docetaxel on PFS would impact on the cost-effectiveness estimates. We showed how such predicted estimate can be obtained and how it can enable implementation of three-state Markov model comprising of stable disease state, separate progressive disease state and death state

Conducting bivariate meta-analysis by modelling jointly correlated treatment effects on OS and PFS enabled us to predict the effect of docetaxel (with prednisolone) on PFS in TAX 327 study which was the only RCT, at the time of technology assessment, evaluating docetaxel and which reported the treatment effect on OS only. To use the predicted effect to inform the three-state model, further analysis was conducted to estimate other parameters, such as transition probabilities between the states and utilities for the stable disease and the progressive disease states, now being the separate states.

The cost effectiveness analysis resulted in much lower ICER obtained from the three-state model compared to the two-state model. This was due to combining stable and progressed patients in the single state in the two state model, which likely underestimated the average utility. The two-state model used utility reported by Sandblom et al[19] who measured EQ-5D twelve months prior to death. Taking into account that the median overall survival is approximately 18 months, 12 months prior to death would be at median six months' time from start of treatment which is equal to the median time to progression. Therefore this measurement of HRQoL corresponds only to patients who progressed. Hence the utility of the StD state in the two state model seems underestimated as some of the patients in the StD state, for a considerable amount of time, are progression free and therefore should have higher utility. In the three-state model we use the same utility in the PD state as in the original two-state model, but allow for the utility in the StD state to vary over time allowing a proportion of patients remaining in the StD state in each cycle to have a higher utility, leading to higher average utility. In addition, most patients leaving the StD state transition to progression and the difference in transition probabilities (for transition out of StD state) between treatments is higher in the three-state model than in the two state model. This is because the HR on PFS is higher than HR on OS. Therefore the difference between treatments in rates of patients leaving the StD state is also higher in the three-state model. This leads to a larger average difference in utility which is relatively larger than the difference in average incremental cost, as substantial care costs are still required post-progression relative to the cost of treatment before progression. This substantial increase in the QALY gained led to the much smaller ICER obtained from the three-state model compared to the two-state model.

There were some limitations of our research. One was the variability in the definitions of the progression endpoints (for details see Appendix A.1 of the supplement). Standardising the definition of the progression endpoint for all the RCTs, however, would require IPD from each of the trials, which is not achievable within the resources of this project. Another limitation was a small number of studies in the bivariate meta-analysis. However, careful analysis of the results indicated that the predicted effect on PFS in Tax 327 had converged in the MCMC simulation. Moreover, the obtained estimate was comparable with the estimate of the effect of docetaxel combination therapy reported in SWOG trial (not included in the meta-analysis due to unlicensed comparator, see Table C2 in the supplement) obtained from the digitised data, but with appropriately larger uncertainty.

In conclusion, we have illustrated that a careful synthesis of all available evidence can help valuable data, otherwise discarded, being used to better inform the decision making process. Multivariate evidence synthesis is, therefore, a valuable tool in synthesis of evidence for medical decision making.

**5. Acknowledgements**

S. Bujkiewicz was supported by the MRC Methodology Research Programme [New Investigator Research Grant, MR/L009854/1]. KR Abrams was supported by the NIHR [Senior Investigator, NF-SI-0512-10159].

**Appendix A: Summary of synthesis of evidence conducted by Collins et al**

As part of the Health Technology Assessment (HTA) (Collins et al., 2007) evaluating the clinical- and cost-effectiveness of docetaxel in combination with either prednisone or prednisolone (D+P) for the treatment of mHRPC, a scoping search for studies evaluating the clinical- and cost-effectiveness of D+P was conducted. As only one RCT was identified to have compared D+P with mitoxantrone plus prednisone (M+P) and no other RCT compared D+P with any other possible interventions, RCTs that assessed mitoxantrone in combination with a corticosteroid compared with any chemotherapy regimen or best supportive care or placebo were also included in the scoping search. Extension of the studies selection to include studies that evaluated mitoxantrone in combination with a corticosteroid was to allow for the comparison between D+P and other relevant interventions using mitoxantrone in combination with a corticosteroid as a common comparator in indirect comparison analysis. In total, seven RCTs were identified based on the inclusion criteria, of which three RCTs used docetaxel compared with M+P, three RCTs used mitoxantrone plus a corticosteroid (M+P/H) compared with a corticosteroid (P/H) and one RCT used M+P compared with mitoxantrone plus prednisone plus clodronate (M+P+Clo). The three RCTs that included docetaxel had docetaxel in the following combination: D+P, docetaxel with estramustine (D+E) and docetaxel with estramustine and prednisone (D+E+P). All studies with details of interventions and reported outcomes are presented in Table A1.

*Table A.1*

| HTA Set | Trial | No. of arms | Reference Treatment | Comparative Treatment(s) | Total no. of patients | OS data | PFS data |
|---|---|---|---|---|---|---|---|
| Set 1* | CCI-NOV22 (Tannock et al., 1996) | 2 | M+P | P | 161 | Yes | Yes |
| | CALGB 9182 (Kantoff et al., 1999) | 2 | M+H | H | 242 | Yes | Yes |
| | Berry et al. (Berry et al., 2002) | 2 | M+P | P | 120 | Yes | Yes |
| | TAX 327 (Tannock et al., 2004) | 3 | M+P | D+P<br>D1+P | 1006 | Yes | No |
| Set 2 | Ernst (Ernst et al., 2003) | 2 | M+P | M+P+Cl | 209 | Yes | Yes |
| | SWOG (Petrylak et al., 2004) | 2 | M+P | D+E | 674 | Yes | Yes |
| | Oudard (Oudard et al., 2005) | 3 | M+P | D70+E+P<br>D35+E+P | 127 | Yes | Yes |

*HTA set in the main manuscript (licensed interventions)
†unlicensed interventions used by Collins et al. in sensitivity analysis (we use SWOG data to obtain some of the transition probabilities).

*A.1 Definitions of outcome measures*

Definitions of OS were consistent for the four RCTs in HTA Set except for trial Berry, where OS was not explicitly defined[9]. Overall survival was defined as the time from the date of randomisation to the date of death or censored at the date when the patient was last known to be alive. There were inconsistencies in the definition of progression across the three RCTs in HTA Set (excluding trial TAX327 which did not report PFS). Specifically, PFS reported for CALGB 9182[10] was defined as the time from the date of randomisation to the date of progression or death, whichever occurred first. Neither PFS nor time to progression (TTP) was reported for trial CCI-NOV22[12]; however, the HTA report presented TTP estimates for this trial. TTP was reported by Berry and colleagues [9] but no explicit definition for TTP was provided.

**Appendix B: Cost-effectiveness analysis methods**

For the cost-effectiveness assessment of D+P in the HTA report, two separate analyses were performed. This is due to the unlicensed status of some of the treatment regimens. The first analysis looked at three interventions (that are licensed at the time of the HTA report submission), namely D+P, M+P(H) and P(H). The second analysis looked at eight interventions, including the three in the first analysis and the following five: D1+P, D+E, D70+E+P, D35+E+P, M+P+Clo. Due to the unlicensed status of the interventions in the second analysis, the economic decision making in this HTA report was based on the results of the first analysis. In the research that follows in this paper, the focus is on the cost-effectiveness assessment of the interventions compared in the first analysis. Data from SWOG trial were used to inform the utility in the cost-effectiveness analysis.

To develop the economic models, data for the construction of transition probabilities, definition of costs and utilities for each of the interventions need to be extracted. These data can be extracted from reviews, single RCT or evidence synthesis from a number of trials/studies. Specifications of the transition probabilities, cost and utilities are described in sections 2.3.1, 2.3.2 and 2.3.3 respectively.

*B.1 Transition probabilities*

For the WinBUGS two-state model, the transition probabilities were estimated using the Weibull parameters reported in the HTA report, which used IPD from trial TAX 327. This was for consistency with the parameters used in the HTA model. As for the three-state model, which incorporated a PD state, the transition probabilities for transition from StD state to PD state were estimated using parametric Weibull survival modelling on reconstructed PFS IPD from one of the six RCTs (excluding trial TAX 327) in the HTA report.

The selection criteria for the RCT to be used for estimating the transition probabilities for treatment arm M+P are: (i) comparable OS profile of the selected trial and trial TAX 327; (ii) selected trial having a mean time of progression closest to the reported mean cycle of M+P administered in trial TAX 327 (that is: 5.9 cycles as reported in the HTA report). As PFS was not recorded for TAX 327, selection criteria (ii) is based on the assumption that the patients in trial TAX 327 M+P arm were administered M+P till progression. The mean number of cycles of M+P administered was then used as a crude approximation of the 'potential' mean time to progression for patients administered M+P in trial TAX 327. Transition probability for transition from the StD state to death state was obtained from an article on cost-effectiveness analysis for advance hormone-dependent prostate cancer (Lu et al., 2012). In the absence of patient-level data on both OS and PFS for any of the interventions, the transition probability for transition from PD state to death state could not be estimated. Methods for the estimation of (i) transition probabilities using parametric Weibull survival model and (ii) transition probabilities for transition from PD state to death state in the three-state model are described in the next two subsections.

### B.1.1 Transition probabilities estimated using parametric Weibull survival model

Survival analysis using the parametric Weibull model was used to implement time-dependency in the transition probabilities in the economic models (transition probabilities for transition from StD state to death state in WinBUGS two-state model; and transition probabilities for transition from StD state to PD state in three-state model). The Weibull distribution takes the following probability density function:

$$f(t) = \lambda \gamma t^{\gamma-1} exp(-\lambda t \gamma)$$

where $\lambda$ gives the scale of the distribution and $\gamma$ defines the shape. The hazard function for this distribution is therefore:

$$h(t) = \lambda \gamma t^{\gamma-1}$$

with a cumulative hazard function of:

$$H(t) = \lambda t^{\gamma}$$

where the survival function is related to the cumulative hazard function in the following form:

$$S(t) = exp[-H(t)]$$

Since hazards are instantaneous, these need to be converted to a transition probability for a given period, such as a Markov cycle. Using the survival function, transition probability between time-points $(t - u)$ and $t$, denoted as $TP(t_u)$ where $u$ is the cycle length, was defined as one minus the ratio of the survival function at the end of the interval to the survival function at the beginning of the interval. This function defined as:

$$TP(t_u) = 1 - S(t)/S(t - u)$$

was re-written using the cumulative hazed function as:

$$TP(t_u) = 1 - exp[-H(t)]/exp[-H(t - u)]$$
$$= 1 - exp[H(t - u) - H(t)]$$

Therefore, transition probability was defined using the Weibull parameters as follows:

$$TP(t_u) = 1 - exp[\lambda(t - u)^\gamma - \lambda t^\gamma]$$

In the HTA report, results of the Weibull survival analysis model were presented in the form of the regression coefficients of the intercept and scale parameters. These two parameters are expressed in terms of the Weibull parameters, λ and γ, as follows:

$$\lambda = exp(-\beta/\alpha)$$

$$\gamma = \frac{1}{\alpha}$$

where $\beta$ is the intercept and $\alpha$ is the scale regression coefficient parameters from the Weibull survival analysis.

When performing the probabilistic analysis, the covariance between the intercept and scale regression parameter from the Weibull survival analysis were also incorporated in the WinBUGS two-state model. This was achieved by using the Cholesky decomposition matrix derived from the covariance matrix obtained from the Weibull survival regression model. Given a covariance matrix of the form:

$$Covariance, C = \begin{pmatrix} a & b \\ b & c \end{pmatrix}$$

the Cholesky decomposition matrix takes the form:

$$D = \begin{pmatrix} \sqrt{a} & 0 \\ \dfrac{b}{\sqrt{a}} & \sqrt{c - \dfrac{b^2}{a}} \end{pmatrix}$$

such that $C = D D^*$ where $D^*$ denotes the conjugate transpose of $D$. Cholesky decomposition matrices of the covariance matrices for the interventions, D+P and M+P, were calculated independently and applied to the transition probabilities of D+P and M+P respectively in the WinBUGS two-state model to allow for the correlation between the intercept and scale parameters when sampling the random normal draws for the two parameters. Assuming that the Cholesky decomposition matrix of the covariance matrix for M+P is:

$$D_{M+P} = \begin{pmatrix} u_{D,M+P} & 0 \\ v_{D,M+P} & w_{D,M+P} \end{pmatrix}$$

the transition probability incorporating parameter uncertainties for transition from StD state to death state for M+P is defined as:

$$TP_{D,M+P}(t_u) = 1 - exp[H_D(t-u) - H_D(t)]$$

$$TP_{D,M+P}(t_u) = 1 - exp[\lambda_{D,M+P}(t-u)^{\gamma_{D,M+P}} - \lambda_{D,M+P} t^{\gamma_{D,M+P}}]$$

where:

$$\lambda_{D,M+P} = \exp\left(\frac{-\beta_{D,M+P}}{\alpha_{D,M+P}}\right)$$

$$\gamma_{D,M+P} = \frac{1}{\alpha_{D,M+P}}$$

and

$$\beta_{D,M+P} = \beta_{M+P} + u_{D,M+P} Z_{\beta,D,M+P}$$

$$\alpha_{D,M+P} = \alpha_{M+P} + v_{D,M+P} Z_{\beta,D,M+P} + w_{D,M+P} Z_{\alpha,D,M+P}$$

where $\beta_{M+P}$ and $\alpha_{M+P}$ are the intercept and scale regression coefficients for M+P presented in the HTA report; and $Z_{\beta,D,M+P} \sim Normal(0,1)$ and $Z_{\alpha,D,M+P} \sim Normal(0,1)$.

Transition probabilities of interventions D+P and M+P for the WinBUGS two-state model were calculated using the regression coefficients from the HTA report (Table 28 (Collins et al., 2007)). Transition probabilities for P were calculated by applying the HR of P versus M+P or

HR of P versus D+P to the hazard rates of M+P and D+P in the transition probabilities respectively. Therefore, assuming that the transition probability for M+P is given by:

$$TP_{M+P}(t_u) = 1 - exp[H(t-u) - H(t)]$$

and with a HR for P versus M+P, denoted as $HR_{P/M+P}$, the transition probability for P is given by:

$$TP_P(t_u) = 1 - exp\{HR_{P/M+P}[H(t-u) - H(t)]\}$$
$$= 1 - exp[H(t-u) - H(t)]^{HR_{P/M+P}}$$

Uncertainty associated with the HR was incorporated in the model by assigning a normal distribution to the logarithm of the HR as follows:

$$LHR \sim Normal(\bar{\mu}, \sigma^2)$$

where $\bar{\mu}$ and $\sigma^2$ are the mean and variance estimate of the LHR from random-effects meta-analysis.

For the three-state model, the set of transition probabilities for intervention M+P was calculated using regression coefficients of the parameters of a Weibull survival model for PFS using re-constructed IPD from one of the RCTs in HTA Full Set selected based on the criteria outlined above. As no PFS patient-level data was available for the interventions D+P and P, transition probabilities for each of the interventions were calculated by applying their HR with respect to M+P to the transition probabilities of M+P. Similarly, uncertainty associated with each of the HRs was included in the respective models by assigning normal distribution to the LHRs.

<u>*B.1.2. Transition probabilities from PD state to death state (three-state model)*</u>

Although IPD were reconstructed for PFS (together with OS) for the trial selected for estimating the transition probability from StD state to PD state, the reconstructed IPD for PFS and OS were not paired by patient. Hence, it would not be possible to estimate the transition probabilities from PD state to death state using parametric survival analysis performed using reconstructed IPD as described in the previous section. To overcome this issue, transition probabilities were estimated by assuming the mean total survival time was equal to the weighted sum of combined survival time from stable disease to progression and then to death and the survival time when death occurred from other causes:

$$\begin{aligned}
mean(Total\ Time) &= W_{StDtoPDtoDeath}[mean(Time_{StDtoPD}) + mean(Time_{PDtoDeath})] \\
&+ W_{StDtoDeath} mean(Time_{StDtoDeath})
\end{aligned}$$

where $mean(Time_{StDtoPD})$ defines the mean time that patients stayed in the StD state before transition to the PD state; $mean(Time_{PDtoDeath})$ and $mean(Time_{StDtoDeath})$ define the mean time for PD state to death state and StD state to death state respectively; W defines the weight assigned to the mean time and is related to the number of patients who transition through the two potential pathways in the economic model as shown in **Error! Reference source not found.** (bottom), from stable disease to death either with or without disease progression.

As the proportion of patients who died of causes unrelated to prostate cancer was expected to be small (<1%), we assumed that $W_{StDtoDeath} \rightarrow 0$, therefore $W_{StDtoPDtoDeath} \rightarrow 1$. Hence,

$$\begin{aligned}
mean(Total\ Time) &= W_{StDtoPDtoDeath}[mean(Time_{StDtoPD}) + mean(Time_{PDtoDeath})] \\
&= mean(Time_{StDtoPD}) + mean(Time_{PDtoDeath})
\end{aligned}$$

and therefore,

$$mean(Time_{PDtoDeath}) = mean(Total\ Time) - mean(Time_{StDtoPD})$$

Assuming that the survival data for patients from PD to death follows an exponential survival distribution, the transition probability between time-points $(t - u)$ and $t$, denoted as $TP(t_u)$ where $u$ is the cycle length, is defined as follows:

$$\begin{aligned}
TP(t_u) &= 1 - exp\{\lambda(t - u)^\gamma - \lambda t^\gamma\} \\
&= 1 - exp(-\lambda u)
\end{aligned}$$

where $\gamma = 1$ for the exponential survival model.

As the hazard rate, $\lambda = \frac{1}{mean(Time)}$,

$$TP(t_u) = 1 - exp\left(\frac{-u}{mean(Time)}\right)$$

For M+P and D+P, the $mean(Total\ Time)$ for each of the interventions were estimated using the mean survival time calculated from the reconstructed OS IPD of trial TAX 327. For P, the mean survival time was estimated by a random-effect meta-analysis of the log hazard rate of the three RCTs that had a P treatment regimen arm.

As PFS endpoint were not recorded for trial TAX 327, the mean number of cycles of drug reported in the HTA report were used to represent the mean time from stable disease to progression, $mean(Time_{StDtoPD})$, based on the assumption that patients stopped drug treatment on the onset of disease progression. Mean number of cycles of drug P was not reported in the HTA report. Therefore, the mean time to progression was also estimated using meta-analysis of the log hazard rate of the two RCTs that reported PFS data for P.

Transition probabilities for transition from PD state to death state for each intervention were therefore calculated using the equation:

$$TP(t_u) = 1 - exp\left(\frac{-u}{[mean(Total\ Time) - mean(Time_{StDtoPD})]}\right)$$

Uncertainty associated with the mean survival time or log hazard rate were also incorporated using normal distributions and propagated in the economic model. As the exponential survival model is a single parameter model, Cholesky decomposition was not required for defining the uncertainty.

### *B.2. Cost*

Cost data comprises drug acquisition and administration cost for each interventions, cost of the management of adverse side effects and subsequent follow up cost that included cost of further chemotherapy after disease progression, management of side-effects and palliative cost. Cost for each of the interventions to be used in the WinBUGS 2-state and 3-state model were extracted from cost data presented in the HTA report. In the report, costs were categorised into three components: namely, (i) the drug cost, (ii) the follow up cost and (iii) the terminal care cost. Drug cost included cost of acquisition and administration of each intervention.

Follow up cost included the cost of managing side-effects, subsequent chemotherapies and hospitalisation for palliative care. Terminal care costs were one-off costs used to incorporate the cost of caring for patients in the last month of life. As stated in the HTA report, terminal care cost data were not recorded in the trial (TAX 327), hence these costs were estimated from patients who died in the first six months after entering the trial. In the absence of costs per cycle for follow up cost, these costs were assigned and calculated as one-off cost, in a similar fashion as terminal care cost, as patient died. Cost data for interventions D+P and M+P were estimated using patient level data from trial TAX 327 while cost data for P were estimated using published cost-effectiveness analyses by Bloomfield and colleagues (Bloomfield et al., 1998).

Gamma distribution was used to represent uncertainty in the follow up costs and terminal care costs. Drug costs for each of the interventions were calculated based on the mean number of cycles of drugs administered. Normal distribution was used to describe the number of cycles of drugs administered to reflect uncertainty in the drug costs.

For the WinBUGS two-state model, the total costs were calculated as the summation of all three categories of costs. For the three-state model, drug costs and terminal care costs were calculated in a similar way to that calculated in the WinBUGS two-state model while follow-up costs were calculated by dividing the follow-up costs into two unequal parts (using a parameter defined as $\psi$ in the equations that follow).

Costs of subsequent chemotherapy and hospitalisations accounted for between 70% and 80% of follow-up costs which most likely occurred post-progression and the remaining follow-up cost (20% to 30%) were related to side effects likely to occur prior to progression (but may also be associated with the subsequent chemotherapy post-progression). Therefore the follow up costs were divided into portions corresponding to StD state and PD state.

As the base case analysis for the three-state model, 75% ($\psi = 0.75$) of the follow-up costs were assigned to the PD state to account for the cost of subsequent chemotherapy, managing side-effects and hospitalisations post-progression. Computation of these costs were based on the number of patients who died per cycle while the remaining 25% of the costs that were assigned to the StD state were computed based on the number of patients who progressed per cycle. Follow-up costs were assigned as one-off cost in a similar way as the WinBUGS two-state model.

In the WinBUGS two-state model, follow-up costs were calculated based on patients died per cycle as:

$$Total\ Cost_{FU} = \sum_{i=1}^{180}\left(Cost_{FU,i} \times N_{Died,i}\right)$$

where $Cost_{FU,i}$ represents follow-up cost data for cycle *i*, and $N_{Died,i}$ represents the number of patients who died in cycle *i*.

For the three-state model, the follow-up costs were calculated as:

$$Total\ Cost_{FU} = Total\ Cost_{StDFU} + Total\ Cost_{PDFU}$$

where:

$$Total\ Cost_{StDFU} = \sum_{i=1}^{180}\left[(1-\psi) \times Cost_{FU,i} \times N_{Progressed,i}\right]$$

$$Total\ Cost_{PDFU} = \sum_{i=1}^{180}\left(\psi \times Cost_{FU,i} \times N_{Died,i}\right)$$

$\psi$ represents the proportion of follow-up costs associated with the PD state (termed "division factor") and $N_{Progressed,i}$ represents the number of patients who progressed in cycle *i*.

Sensitivity analyses for assessing different proportions of follow-up costs associated with the PD state are described in Section **Error! Reference source not found.**. An annual discount rate of 3.5% was used for discounting the cost after the first year.

### B.3 Utility

Utility for StD state in the WinBUGS two-state model was defined using a beta distribution with parameter values Beta(21.1, 18.1), derived from a mean EQ-5D HRQoL of 0.538 (95% CI: 0.461 to 0.615) as reported in the study by Sandblom and colleagues (Sandblom et al., 2004). For the three-state model, the utility distribution, Beta(21.1, 18.1), for the StD in the two-state model was assigned as the utility distribution for the PD state as the corresponding mean EQ-5D of 0.538 represents the HRQoL of all patients 12 months prior to death. Two additional EQ-5D values as discussed in Section 2.3.3 were extracted from the Sandblom study (Sandblom et al., 2004), they were the EQ-5D for patients who were still surviving at the time of analysis of the study, EQ-5D = 0.770 (95% CI: 0.755 to 0.785), and EQ-5D of patients who died of other non-prostate cancer related death, EQ-5D = 0.564 (95% CI: 0.497 to 0.631). These two utility data were defined in the economic model using the following beta distributions, Beta(581.3, 173.6) and Beta(29.1, 22.5) respectively. Utility for the StD state was calculated based on the method described in Section 2.3.3 using two additional EQ-5D values extracted from the Sandblom study (Sandblom et al., 2004), they were the EQ-5D for patients who were still surviving at the time of analysis of the study, EQ-5D = 0.770 (95% CI: 0.755 to 0.785), and EQ-5D of patients who died of other non-prostate cancer related death, EQ-5D = 0.564 (95% CI: 0.497 to 0.631). These two utility data were defined in the economic model using the following beta distributions, Beta(581.3, 173.6) and Beta(29.1, 22.5) respectively to derive the utility for the StD state.

As the utilities defined were selected to reflect the general HRQoL of patients with advanced prostate cancer and were independent of the interventions administrated by the patients, the utilities were used in the model for all three interventions.

### B.4. Cost effectiveness

Cost effectiveness of interventions was assessed by obtaining the incremental cost-effectiveness ration (ICER). ICER was calculated by taking the difference between the mean values of the cost of interventions over the difference between the mean values of the QALYs gained of interventions; as:

$$ICER = \frac{\overline{Cost_{D+P}} - \overline{Cost_{M+P}}}{\overline{QALY_{D+P}} - \overline{QALY_{M+P}}}$$

where $\overline{Cost_{D+P}}$ and $\overline{Cost_{M+P}}$ defines the mean cost of D+P and M+P respectively and $\overline{QALY_{D+P}}$ and $\overline{QALY_{M+P}}$ defines the mean QALY gained per patient for D+P and M+P respectively.

**Appendix C: Additional results**

**C.1 Re-constructed IPD summary statistics**

Tables A1 and A2 show HRs for OS and PFS respectively, for all seven studies listed in Table A1 for completeness.

*Table C1:Individual trial's HRs on OS obtained using IPD reconstructed from Kaplan-Meier survival curves*

| Trial | Comparison | HR (95% CI) reported in journal article | HR (95% CI) reported in HTA report | HR (95% CrI) from reconstructed IPD |
|---|---|---|---|---|
| *Overall Survival* | | | | |
| TAX 327 | D+P / M+P | 0.76 (0.62-0.94) | 0.76 (0.62-0.94) | 0.76 (0.620, 0.936) |
| CALGB 9182 | M+H / H | Not reported but median survival reported as: M+H 12.3 months and; H 12.6 months (p=0.77) | 1.05 (0.74, 1.49) | 0.96 (0.732, 1.251) |
| CCI-NOV 22 | M+P / P | Not reported but a total of 140 deaths reported at time of analysis (p=0.27) | 0.91 (0.69, 1.19) | 0.81 (0.590, 1.110) |
| Berry | M+P / P | Not reported but median survival reported as: M+P 23 months and; P 19 months (p=0.569) | 1.13 ( 0.75, 1.70) | 0.95 (0.628, 1.432) |
| Ernst | M+P+Cl/M+P | 1.05 (0.78, 1.42) | 1.05 (0.78, 1.42) | 1.08 (0.799, 1.452) |
| SWOG | D+E / M+P | 0.8 (0.67, 0.97) | 0.8 (0.67, 0.97) | 0.79 (0.659, 0.955) |
| Oudard | D70+E+P / M+P | Not reported but median survival reported as: D70+E+P 18.6 months, D35+E+P 18.4 months and; M+P 13.4 months | 0.94 (0.29, 1.02) | 1.08 (0.675, 1.715) |
| | D35+E+P / M+P | | 0.86 (0.68, 1.08) | 0.75 (0.448, 1.245) |

*Table C2: Individual trial's HRs on PFS obtained using IPD reconstructed from Kaplan-Meier survival curves*

| Trial | Comparison | HR (95% CI) reported in journal article | HR (95% CI) reported in HTA report | HR (95% CrI) from reconstructed IPD |
|---|---|---|---|---|
| *Progression-free Survival* | | | | |
| TAX 327 | D+P / M+P | Endpoint not collected | Not possible | Not possible |
| CALGB 9182 | M+H / H | Not reported but median survival reported as: M+H 3.7 months and; H 2.3 months (p=0.0218) | Time to progression (calculated from number of events and p-value presented in the trial publication) HR= 1.50 (1.06, 2.13); p = 0.0218 | 0.74 (0.574, 0.954) |
| CCI-NOV 22* | M+P / P | Not reported | 0.47 (0.32, 0.68) | Not possible+ |
| Berry* | M+P / P | Not reported but median survival reported as: M+P 8.1 months and; P 4.1 months (p=0.018) | Estimated from the Kaplan-Meier curve for PFS presented in the trial publication. HR= 0.64 (0.48, 0.86) | 0.63 (0.432, 0.927) |
| Ernst | M+P+Cl/M+P | 0.81 (0.61, 1.07) | 0.81 (0.61, 1.07) | 0.84 (0.63, 1.112) |
| SWOG | D+E / M+P | Not reported but median survival reported as: D+E 6.3 months and; M+P 3.2 months (p<0.001) | time to disease progression observed for the docetaxel group compared with the mitoxantrone group: HR=1.30 (1.11, 1.52); p < 0.001 | 0.73 (0.627, 0.860) |
| Oudard* | D70+E+P / M+P D735+E+P / M+P | Not reported but median survival for time to PSA progression is reported as: D70+E+P 8.8 months, D35+E+P 9.3 months and; M+P 1.7 months | Not reported Not reported | Not possible Not possible |

*Trials where TTP was reported in the journal article or HTA report instead of PFS
+No Kaplan-Meier survival curve in published article

**C.2 Justification for choice of Trial SWOG**

Overall survival curves for four RCTs in HTA Full Set (excluding CALGB 9182 which used hydrocortisone instead of prednisone and CCI-NOV22 which did not report PFS) were compared to the OS curve of trial TAX 327 and presented in Figure. The OS Kaplan-Meier curves suggested that trial SWOG has an OS profile closest to trial TAX 327.

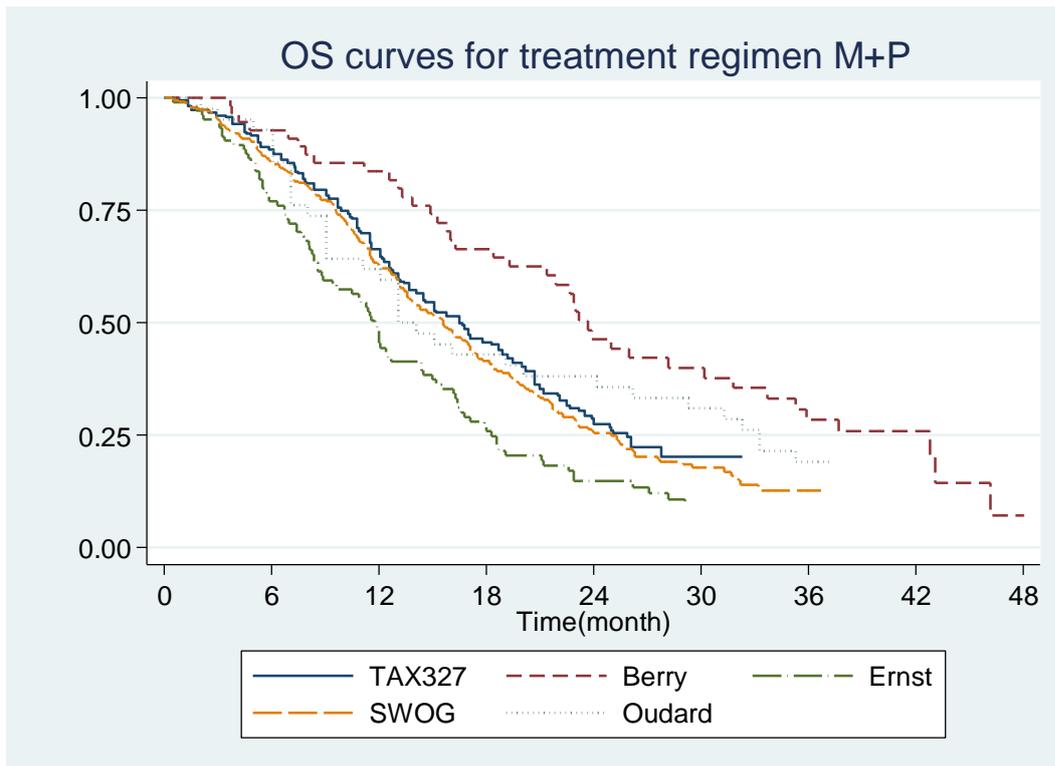

*Figure C1: Overall survival Kaplan-Meier curves for RCTs in the HTA report*

Mean time to progression for patients administered M+P for the four RCTs were estimated using the IPD reconstructed from PFS Kaplan-Meier curves. As described in Section 2.3.1, as PFS was not recorded in trial TAX 327, mean number of cycles of M+P administered in trial TAX 327 was used for comparing with the mean time to progression in the four RCTs. The mean time to progression for the four RCTs and the mean number of cycles of M+P administered in trial TAX 327 are shown in Table. Trials Ernst and SWOG have mean time to progression closest to the assumed mean time to progression of TAX 327. The results suggested that trial SWOG has an OS profile closest to trial TAX 327 and therefore potentially a PFS profile closest to TAX 327 if the PFS endpoint had been recorded. Hence, reconstructed IPD of trial SWOG were used to estimate the transition probabilities from StD state to PD state for M+P.

*Table C3: Mean time to progression for RCTs in the HTA report*

| Trial | Mean Time to Progression (SE) |
|---|---|
| Berry et al. | 12.8 (1.63) |
| Ernst | 5.9 (0.53) |
| SWOG | 5.9 (0.33) |
| Oudard | 4.2 (0.88) |
| | Mean no.of cycles (SE) |
| TAX 327 | 5.9 (0.17) |

**C.3 Cost and utility**

Costs of drug for interventions M+P and D+P were estimated using the mean cycle of chemotherapy administrated. The reported mean number of cycles for the interventions M+P and D+P were 5.9 (SE=0.17) and 7.3 (SE=0.18) respectively, Normal distributions were assigned to the number of treatment cycles to incorporate uncertainty around these values in the two economic models. Cost of drug for M+P was £347.73/cycle and £1253.92/cycle for D+P, including £177.46/cycle for outpatient attendance fees for both drug regimens. Cost of drug for P was calculated at £1.48 per patient per cycle. In the WinBUGS two-state model, the drug costs for a patient taking P was calculated for the number of cycles that the patient remains in the StD state before transition to the death state. In the three-state model, the cost drug for a patient taking P was calculated for the number of cycles that the patient remains in the StD state before transition to the PD state. It was assumed that the patient would stop taking P after progression and hence, no drug cost would be calculated for the cycles post-progression before transition to the death state.

Follow-up and terminal care costs for interventions M+P and D+P were estimated from trial TAX 327 as reported in the HTA report (Table 36 and Table 37 in (Collins et al., 2007)). Uncertainties for the costs were applied in the two economic models using Gamma distributions. Follow-up and terminal care costs for drug P were not available and were estimated from the costs of intervention M+P from trial TAX 327. In order to estimate the costs for drug P, a cost ratio of drug P with reference to drug M+P was estimated using costing data of P and M+P from a review article (Bloomfield et al., 1998). The mean cost ratio estimated in the WinBUGS two-state model was 1.278 (95% CrI: 0.946 to 1.691) which suggested that the mean cost of P was higher than the mean cost of M+P. This cost ratio was calculated by assigning Gamma distributions [Gamma($\alpha,\beta$)] of Gamma(105, 276) and Gamma(81, 285) to the cost data of P and M+P respectively. The mean cost (drug, follow-up and terminal care) per patient at each state in the economic model for each of the interventions are presented in Table C4.

Mean EQ-5D HRQoL for all patients in the 12 months prior to death was 0.538 (95% CI: 0.461 to 0.615) as reported in the study by Sandblom and colleagues (Sandblom et al., 2004). Using this EQ-5D data, the utility for StD state in the WinBUGS two-state model was defined using a beta distribution with parameter values: Beta(21.1, 18.1). For the three-state model, the utility distribution, Beta(21.1, 18.1), for the StD in the WinBUGS two-state model was assigned as the utility distribution for the PD state. Two additional EQ-5D values as discussed in Section 2.3.3 were extracted from the Sandblom study (Sandblom et al., 2004), they were the EQ-5D for patients who were still surviving at the time of analysis of the study, EQ-5D = 0.770 (95% CI: 0.755 to 0.785), and EQ-5D of patients who died of other non-prostate cancer related death, EQ-5D = 0.564 (95% CI: 0.497 to 0.631). These two utility data were defined in the economic model using the following beta distributions, Beta(581.3, 173.6) and Beta(29.1, 22.5) respectively. Utility for the StD state was calculated based on the method described in Section 2.3.3.

As the utilities defined were selected to reflect the general HRQoL of patients with advanced prostate cancer and were independent of the interventions administered by the patients, the utilities were used in the model for all three interventions. Mean QALY per patient for each of the interventions in the economic models are presented in Table C4.

*Table C4: Mean cost, mean QALY and mean time spent per patient at each state in the economic model'*

| Economic Model | Drug | Mean Cost (£) (95% CrI) | Drug cost (£) | Mean QALYs (95% CrI) | Mean Time (95% CrI) |
|---|---|---|---|---|---|
| *WinBUGS two-state model* | | | | | |
| Stable & Progression Disease State | P (direct) | 11772 (6127, 20280) | 26 (23, 31) | 0.809 (0.5590, 1.0760) | 18.1 (15.54, 21.03) |
| | P (indirect) | 11772 (6128, 20290) | 27 (22, 32) | 0.812 (0.5476, 1.1100) | 18.2 (14.68, 22.41) |
| | M+P | 11237 (6855, 17030) | 2057 (427, 3679) | 0.813 (0.5718, 1.0580) | 18.2 (16.55, 19.93) |
| | D+P | 15862 (9066, 23020) | 9152 (3261, 15050) | 0.967 (0.6746, 1.2690) | 21.9 (19.50, 24.58) |
| *WinBUGS three-state model* | | | | | |
| Stable Disease State | P (direct) | NA | 6 (1, 18) mean cycles = 4.12 (SE:1.47) | 0.276 (0.0614, 0.7646) | 4.1 (0.68, 12.03) |
| Progression Disease State | P (direct) | NA | | 0.573 (0.2667, 1.0080) | 13.4 (6.63, 22.81) |
| | | 10152 (5160.0, 17760.0)[#] | | 0.849 (0.4389, 1.4400) | 17.5 (9.32, 28.90) |
| Stable Disease State | M+P | 371 (211.6, 589.9) | 2047 (400, 3678) mean cycles = 5.34 (SE:0.17) | 0.377 (0.3330, 0.4247) | 5.7 (5.05, 6.40) |
| Progression Disease State | M+P | 3804 (2088.0, 6345.0) | | 0.512 (0.3596, 0.6674) | 11.9 (10.78, 13.12) |
| Terminal care | M+P | 3756 (1026.0, 8239.0) | | 0.889 (0.7200, 1.0600) | 17.6 (16.32, 18.99) |
| | | 9977 (5995.0, 15250.0) | | | |
| Stable Disease State | D+P | 373 (218.1, 578.7) | 9164 (3268, 15000) mean cycles = 6.62 (SE:0.26) | 0.619 (0.4967, 0.7602) | 9.6 (7.67, 11.93) |
| Progression Disease State | D+P | 2474 (1370.0, 4009.0) | | 0.522 (0.3664, 0.6796) | 12.3 (11.15, 13.55) |
| Terminal care | D+P | 3326 (916.2, 7266.0) | | 1.141 (0.9369, 1.3520) | 22.0 (19.75, 24.40) |
| | | 15337 (8594.9, 22440.0) | | | |

#Calculated using mean cost ratio of 1.278 (95% CrI: 0.946 to 1.691) for P compared to M+P